\documentclass[letterpaper]{article} % DO NOT CHANGE THIS
\usepackage{aaai25}  % DO NOT CHANGE THIS
\usepackage{times}  % DO NOT CHANGE THIS
\usepackage{helvet}  % DO NOT CHANGE THIS
\usepackage{courier}  % DO NOT CHANGE THIS
\usepackage[hyphens]{url}  % DO NOT CHANGE THIS
\usepackage{graphicx} % DO NOT CHANGE THIS
\usepackage{natbib}  % DO NOT CHANGE THIS AND DO NOT ADD ANY OPTIONS TO IT
\usepackage{caption} % DO NOT CHANGE THIS AND DO NOT ADD ANY OPTIONS TO IT
\usepackage{algorithm}
\usepackage{algorithmic}
\usepackage{amsmath}
\usepackage{booktabs} 
\usepackage{multirow} 
\usepackage{makecell}

\usepackage{newfloat}
\usepackage{listings}
\DeclareCaptionStyle{ruled}{labelfont=normalfont,labelsep=colon,strut=off} % DO NOT CHANGE THIS
\floatstyle{ruled}
\newfloat{listing}{tb}{lst}{}
\floatname{listing}{Listing}
\title{Enhancing Audiovisual Speech Recognition \\ through Bifocal Preference Optimization}
\author{
    Yihan Wu\textsuperscript{\rm 2,\rm 1},
    Yichen Lu\textsuperscript{\rm 1},
    Yifan Peng\textsuperscript{\rm 1},
    Xihua Wang\textsuperscript{\rm 2},
    Ruihua Song\textsuperscript{\rm 2}\thanks{Corresponding authors: Ruihua Song (rsong@ruc.edu.cn) and Shinji Watanabe (swatanab@cmu.edu).},
    Shinji Watanabe\textsuperscript{\rm 1}\footnotemark[1]
}
\affiliations {
    \textsuperscript{\rm 1}Carnegie Mellon University,
    \textsuperscript{\rm 2}Gaoling School of Artificial Intelligence, Renmin University of China, \\
    \{yihanwu, xihuaw, rsong\}@ruc.edu.cn, \{yichenl5, yifanpen, swatanab\}@andrew.cmu.edu
}

\usepackage{bibentry}
\begin{document}

\maketitle

\begin{abstract}
Audiovisual Automatic Speech Recognition (AV-ASR) aims to improve speech recognition accuracy by leveraging visual signals. It is particularly challenging in unconstrained real-world scenarios across various domains due to noisy acoustic environments, spontaneous speech, and the uncertain use of visual information. Most previous works fine-tune audio-only ASR models on audiovisual datasets, optimizing them for conventional ASR objectives. However, they often neglect visual features and common errors in unconstrained video scenarios. In this paper, we propose using a preference optimization strategy to improve speech recognition accuracy for real-world videos. First, we create preference data via simulating common errors that occurred in AV-ASR from two focals: manipulating the audio or vision input and rewriting the output transcript. Second, we propose \textbf{BPO-AVASR}, a \textbf{B}ifocal \textbf{P}reference \textbf{O}ptimization method to improve \textbf{AV-ASR} models by leveraging both input-side and output-side preference. Extensive experiments demonstrate that our approach significantly improves speech recognition accuracy across various domains, outperforming previous state-of-the-art models on real-world video speech recognition\footnote{We will release code and data in https://github.com/espnet/espnet.}.
\end{abstract}

\section{Introduction}

\label{sec:intro}
% background
Recently, there has been a growing demand for Automatic Speech Recognition (ASR) systems to evolve into unconstrained audiovisual scenarios, such as YouTube videos, online meetings, and live broadcasts. With both audio and visual streams as input, visual information may help improve speech recognition accuracy, especially in cases where the audio is noisy or unclear. There are usually two scenarios for AudioVisual ASR (AV-ASR), focusing solely on lip motion~\cite{DeepAV2022Afouras,E2eAV2021Ma,Lip2019Chung} and using full-frame visual features. In this work, we focus on the latter scenario, improving the AV-ASR performance in unconstrained real-world video scenarios. In such settings, the entire visual frame may contribute to ASR performance by providing additional cues on specific objects, background location, or context.

\begin{figure}[t]
\centering
\includegraphics[width=0.9\linewidth]{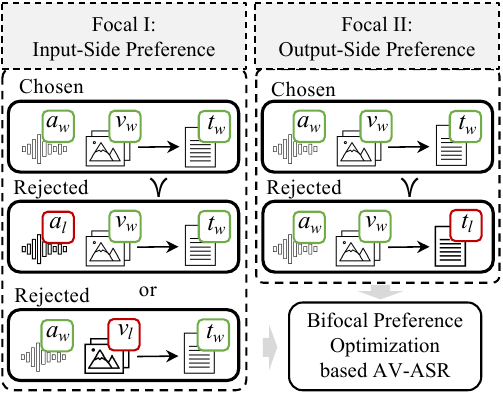} 
\caption{\textbf{Overview of our proposed framework.} We construct bifocal preferences by augmenting transcript (T), audio (A) or video (V), resulting in two types of preference pairs: input-side (A and V) and output-side (T).
Through bifocal preference optimization, BPO-AVASR learns to generate transcripts that are better aligned with these preferences.}

\label{fig:model_arch}

\end{figure}

Compared to standard ASR, performing AV-ASR on real-world videos presents the following challenges:
\noindent
\begin{itemize}
    \item \textbf{Noisy Acoustic Environments.} Homophone issues pose a significant challenge for both standard ASR and AV-ASR. Real-world recordings with variable acoustic environments further severe homophone problems.
     \item \textbf{Spontaneous Speech Scenarios.} Real-world videos often contain spontaneous conversations, which are more variable than read speech or lectures, making accurate recognition more challenging.
    \item \textbf{Uncertain Use of Vision.} The diversity of video domains and unconstrained scenarios leads to the uncertain use of visual information for AV-ASR systems. While some visual cues are irrelevant, others provide crucial localized details or global context for understanding speech, making it challenging to effectively leverage visual information for improved recognition accuracy.
\end{itemize}
Most current AV-ASR works~\cite{How22018Sanabria,LLD2021Ghorbani,VAT2019Caglayan,AVFormer2023Seo} build upon ASR models by incorporating visual features as conditions. These models typically adhere to the original ASR training objective, primarily involving the next-token prediction tasks based on the original audiovisual data. 
However, these supervised fine-tuning (SFT) based methods do not specifically optimize for visual inputs or address scenarios like noisy acoustic environments and spontaneous conversations, which are common in unconstrained AV-ASR tasks. Consequently, this approach may lead the AV-ASR model to neglect visual features or struggle to adapt to real-world videos that feature noisy and spontaneous speech.

To overcome the aforementioned challenges, we propose a Bifocal Preference Optimization based AV-ASR (BPO-AVASR) framework, which aims to facilitate the development of AV-ASR models with stronger speech recognition ability in unconstrained real-world video scenarios (as shown in Figure~\ref{fig:model_arch}). 
Specifically, we propose a bifocal preference optimization strategy, where one bifocal point is input-side preference and the other focal point is output-side preference. 
Accordingly, we construct the bifocal preference dataset by simulating common errors associated with the above challenges, including homophone errors, poor performance on spontaneous speech, and inadequate use of visual cues. Subsequently, we leverage these pairs to optimize the AV-ASR model's preferences. By emphasizing the generation of correct transcripts considering unconstrained audiovisual inputs, BPO-AVASR can better adapt to recognizing speech for real-world videos.
The main contributions of this work are summarized as follows:

\begin{itemize}
    \item We introduce \textbf{BPO-AVASR}, a novel framework with \textbf{B}ifocal \textbf{P}reference \textbf{O}ptimization specifically to align audio-only ASR models with AV-ASR tasks, optimizing AV-ASR as a preference alignment problem.
    \item We design bifocal preference optimization, a training strategy that optimizes preference for AV-ASR tasks considering both input-side preference and output-side preference. Specifically, we construct the preference data pairs to simulate the common errors in unconstrained video AV-ASR, creating negative samples by augmenting audio, visual, and transcript information.
    \item Through extensive experiments, we demonstrate the effectiveness of our approach, showing a notable enhancement compared with SFT-based models. Additionally, BPO-AVASR achieves superior performance compared to previous state-of-the-art models across three datasets.
\end{itemize}

\begin{figure*}[t]
\centering
\includegraphics[width=0.89\linewidth]{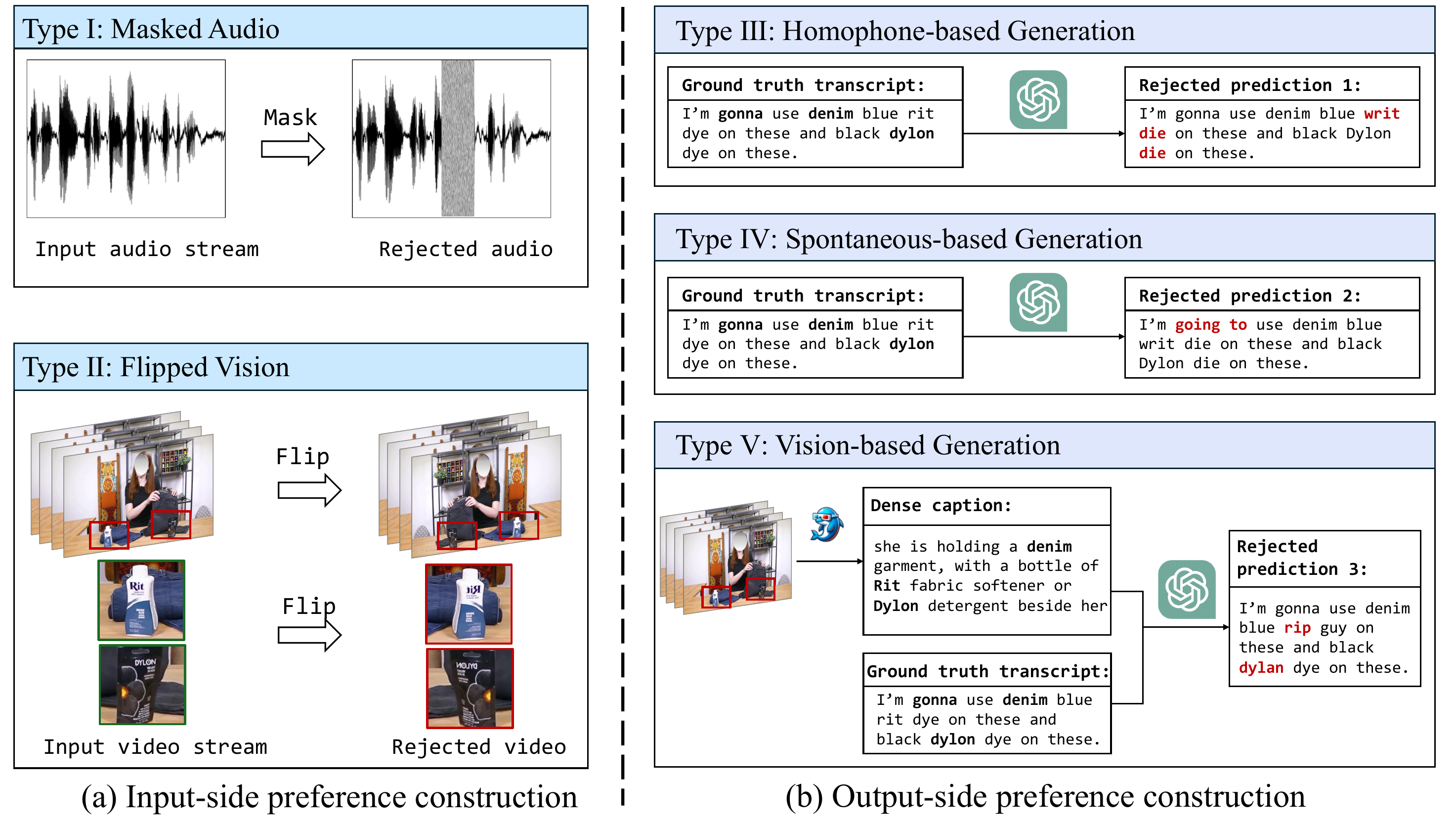} 

\caption{\textbf{The illustration of preference dataset constructing strategies:} (a) Input-side preference construction: simulates errors by manipulating input audios and videos; (b) Output-side preference construction: simulates common errors by manipulating ground truth transcripts.}
\label{fig:data_construct}

\end{figure*}

\section{Related Works}
\subsection{Audiovisual Speech Recognition}
Recent state-of-the-art ASR models have achieved impressive performance on audio-only benchmarks~\cite{ASRsurvey2021Li,E2Esurvey2024Prabhavalkar}. Whisper-style ASR models~\cite{whisper2023Radford,OWSM2024Peng}, in particular, leverage large-scale supervised learning to achieve robust results across various benchmarks. Building on pre-trained ASR models, several studies explore ASR to audiovisual scenarios~\cite{AVATAR2022Gabeur,Promptwhisper2023Peng,Lip2019Chung, LLD2021Ghorbani}. Most previous AV-ASR works focus on lip motion~\cite{Lip2019Chung,DeepAV2022Afouras,E2eAV2021Ma}.
Moreover, to explore unconstrained AV-ASR for real-world videos, recent works explore different adaptation methods to fine-tune or retrain ASR models with full-frame visual features~\cite{MultiRes2017Paraskevopoulos,AVATAR2022Gabeur,LLD2021Ghorbani,Promptwhisper2023Peng,VisualER2023Kumar,syneslm2024lu}. AVFormer~\cite{AVFormer2023Seo} shows the state-of-the-art performance by integrating visual information into a frozen ASR model and fine-tuning it on the large-scale audiovisual dataset HowTo100M~\cite{HowTo1002019Miech}. 
However, these models are all trained or supervised fine-tuned toward standard ASR optimization objectives, leading to modality inefficient utilization of audio and visual information, particularly in unconstrained real-world video scenarios. In contrast, this work introduces a novel approach designed to adapt ASR models to unconstrained AV-ASR using bifocal preference optimization.

\subsection{Direct Preference Optimization}
To align a pre-trained Large Language Model (LLM) with specific preferences of downstream tasks, Direct Preference Optimization (DPO)~\cite{dpo2023Rafailov} is proposed to optimize the LLM with a single-stage policy learning. DPO demonstrates strong performance while eliminating the need for a separate reward model. 
Intuitively, given the input $x$ and the outputs $y_{w}$ and $y_{l}$\footnote{$w$ and $l$ refer to win and loss respectively.}, DPO maximizes the difference between the reward for the chosen output $r(x, y_{w})$ and that for the rejected output $r(x, y_{l})$. Specifically, given a policy model to be optimized $\pi_{\theta}$ and a reference model $\pi_{\text{ref}}$, DPO formulates the reward as:
\begin{equation}
\label{eq:dpo_reward}
    r(x, y) = \beta \log \frac{\pi_{\theta}(y|x)}{\pi_{\text{ref}}(y|x)} + Z(x),
\end{equation}
where $Z(x)$ is a partition function, and $\beta$ is the hyper-parameter that controls the deviation from the reference model. Based on the Bradley-Terry model~\cite{bradley1952rank}, the optimization objective of DPO becomes
\begin{equation}
\label{eq:dpo_loss_1}
\resizebox{0.4\textwidth}{!}{$
\begin{split}
    \mathcal{L}_{\text{DPO}} \! = \! -\log \sigma \left( \beta \log \frac{\pi_{\theta}(y_w|x)}{\pi_{\text{ref}}(y_w|x)} \! - \! \beta \log \frac{\pi_{\theta}(y_l|x)}{\pi_{\text{ref}}(y_l|x)} \right),
\end{split}
$}
\end{equation}
where $\sigma$ is the logistic function.
Furthermore, the subsequent work of DPO~\cite{DPO22024Rafailov} proves that DPO implicitly learns a token-level reward function, highlighting its potential of using DPO in token-level preference optimization, such as ASR and machine translation~\cite{Tim2024zeng}.

In multimodal scenarios, recent works mainly focus on constructing multimodal preference data to support DPO training in the visual language domain~\cite{mdpo2024Wang,zhang2024directpre,SeVa2024zhu}. 
MDPO~\cite{mdpo2024Wang} designs a conditional preference optimization strategy to improve multimodal preference alignment in vision language scenarios, while SeVa~\cite{SeVa2024zhu} constructs preference datasets via self-supervised learning. 
In this work, to the best of our knowledge, we first introduce DPO to AV-ASR tasks, showing a significant improvement over SFT-based AV-ASR models. 
Also, we leverage bifocal preference optimization, specifically designed for unconstrained AV-ASR tasks. Unlike DPO, which constructs negative samples by focusing solely on the output-side, BPO optimizes the AV-ASR model using both input-side preferences and output-side preferences.

\section{Proposed Method}
\label{sec:method}
We describe our proposed BPO-AVASR framework in this section. To support bifocal alignment optimization, we design two key strategies to construct the bifocal preference dataset, constructing both input-side preferences and output-side preferences (Section~\ref{sec:dataset_construction}).
Then we outline the training procedure, including the supervised fine-tuning stage (Section~\ref{sec:SFT}) and the preference optimization stage (Section~\ref{sec:preference_opt}).

\subsection{Creation of Bifocal Preference Dataset}
\label{sec:dataset_construction}
Suppose we have an audiovisual speech recognition dataset $\mathcal{D} = \{(a^{i}, v^{i}, t^{i})\}_{i=1}^{K}$, which contains $K$ elements. Each element $\tau^{i} = (a^{i}, v^{i}, t^{i})$ consists of a speech $a^{i}$, a video $v^{i}$, and the corresponding transcript $t^{i}$. For AV-ASR tasks, both $a^{i}$ and $v^{i}$ served as the inputs (denoted collectively as $c^{i}$), while the corresponding transcript $t^{i}$ is the output. 
Thus each element can be better denoted as $\tau^{i} = (t^{i} | c^{i}) = (t^{i} | a^{i}, v^{i})$. 
Given the challenges of performing AV-ASR on real-world videos (as analyzed in Section~\ref{sec:intro}), we construct preference pairs by manipulating three modalities, resulting in input-side preferences (audio, video) and output-side preferences (text). By simulating common errors in AV-ASR and constructing the preference dataset accordingly, the model can be optimized by maximizing the distinction between chosen and rejected pairs. This approach enables the model to learn to avoid these kinds of errors during inference.

\paragraph{Input-side Preference.}
To simulate insufficient use of input information, we create rejected samples by independently manipulating the audio and video inputs.
\begin{itemize}
    \item \textbf{Masked Audio.} We mask certain audio frames to simulate situations where speech is noisy or ambiguous. In detail, we mask content words and add Gaussian noise to the corresponding audio input, as shown in Type \uppercase\expandafter{\romannumeral1} of Figure~\ref{fig:data_construct}. Therefore, we obtain rejected sample $\tau_{l} = (t_w^{i} | c_l^{i})= (t_w^{i} | a_l^{i}, v_w^{i})$, which serves as a hard negative example to a chosen sample $ \tau_{w} = (t_w^{i} | a_w^{i}, v_w^{i})$.
    \item \textbf{Flipped Vision.} To mimic the situation where the AV-ASR model uses visual information inadequately, we intentionally construct samples with rejected visual inputs. As shown in Type \uppercase\expandafter{\romannumeral2} of Figure~\ref{fig:data_construct}, we create the rejected sample $v_l^{i}$ by flipping the video $v_w^{i}$ to simulate the situation where the ASR model lacks sufficient visual information, especially for the detailed object information. Flipped images can make optical character recognition (OCR) harder and induce potential rejected responses. Therefore, we obtain the rejected sample $\tau_{l} = (t_w^{i} | c_l^{i}) = (t_w^{i} | a_w^{i}, v_l^{i})$  for a chosen sample $ \tau_{w} = (t_w^{i} | a_w^{i}, v_w^{i})$.
\end{itemize}
As a result, we obtain the preference dataset $\widetilde{D}_{c}$.

\paragraph{Output-side Preference.}
To construct a preference dataset that includes rejected samples simulating common errors in recognized text, we use ChatGPT as a proxy for the human generator. This approach allows us to build a cost-effective yet efficient output-side preference dataset. More specifically, for each ground truth element $\tau_{w}= (t^{i}_{w} | c^{i}_{w}) = (t^{i}_{w} | a^{i}_{w}, v^{i}_{w})$, we generate a rejected text $t^{i}_{l}$ to construct the corresponding rejected element $\tau_{l} = (t^{i}_{l} | c^{i}_{w}) = (t^{i}_{l} | a^{i}_{w}, v^{i}_{w})$.
To account for the different causes of text recognition errors, we employ three distinct prompts for the ChatGPT, each designed to generate rejected text based on specific error types. 
\begin{itemize}
    \item \textbf{Homophone-based Generation.} Given the ground truth text $t^{i}_{w}$, we prompt ChatGPT to replace the word with their homophone, e.g., ``dye'' is replaced by ``die'' (See Type \uppercase\expandafter{\romannumeral3} of Figure~\ref{fig:data_construct}). We specifically focus on content words, as ASR errors frequently involve homophone and near-homophone errors in these words.
    \item \textbf{Spontaneous-based Generation.} For real-world videos, in addition to read speech, there is a significant amount of spontaneous speech. 
    To address this error, we use ChatGPT to generate rejected text $t^{i}_{l}$ from the ground truth text $t^{i}_{w}$. It simulates typical ASR errors in spontaneous scenarios, such as generating ``going to'' instead of ``gonna'' (See Type \uppercase\expandafter{\romannumeral4} of Figure~\ref{fig:data_construct}).
    \item \textbf{Vision-based Generation.} Neglecting or misusing visual information is another common error in AV-ASR. To simulate errors where visual information is ignored during recognition, we remove parts of the ground truth text that are related to the video. Specifically, we transfer the video stream $v^{i}_{w}$ into dense and high-quality caption via ShareGPT4Video~\cite{chen2024sharegpt4video}. Then, we use this caption and ground truth text $t^{i}_{w}$ as a prompt to replace the words that appear in the dense caption, simulating the omission of visual cues.
    For example, ``Rit'' in the frame is re-written as ``rip'' (See Type \uppercase\expandafter{\romannumeral5} of Figure~\ref{fig:data_construct}).
    Rather than using video directly, we observe that using these detailed video captions allows for better control in the construction process.
\end{itemize}
As a result, we obtain the output-side preference dataset $\widetilde{D}_{p}$ by rewriting the transcripts systematically.

\subsection{Supervised Fine-tuning}
\label{sec:SFT}
\paragraph{Pre-trained ASR Model.}
\label{sec:owsm}
We use the audio-only ASR model OWSM v3.1\footnote{We
opt for OWSM over Whisper due to potential data
contamination concerns. Whisper’s training data, sourced
through web crawling, might include our test set.} as our backbone~\cite{OWSM2024Peng}. OWSM v3.1 is
an open-source pre-trained speech recognition model that achieves robust performance on standard ASR benchmarks~\cite{LibriSpeech2015Panayotov,Aishell2017Bu,MLS2020Pratap,tedlium2018Hernandez}. It utilizes an encoder-decoder architecture~\cite{Transformer2017Vaswani}, with the stack of E-Branchformer encoders~\cite{Ebranchformer2022Kim} and Transformer decoders~\cite{Transformer2017Vaswani}.
Being trained on large amounts of ASR data, OWSM v3.1 has a good generalization ability for AV-ASR tasks.

\paragraph{Visual Encoder.}
We leverage a pre-trained visual encoder to extract visual tokens as conditions to the audio-only ASR model. Specifically, we use CLIP~\cite{clip2021Radford} to ensure that all relevant information within the video frames is effectively encoded for AV-ASR.  
We sample $M$ frames from video $v^{i}$ uniformly, then use the CLIP to extract visual tokens from these frames. A visual projection layer is subsequently applied to map visual tokens into speech space.

\paragraph{Fine-tuning Objective}
Given $M$ visual tokens and $N$ speech tokens, we concatenate them along the sequence dimension, resulting in a combined sequence of length $N + M$. 
Then, we feed them into the ASR model and fine-tune the model $\pi_{\text{ref}}$ in Equation~\ref{eq:dpo_reward} with attention loss $\mathcal{L}_{\text{ATT}}$ and the CTC loss $\mathcal{L}_{\text{CTC}}$:
\begin{equation}
\begin{split}
    \label{eq:SFT_loss}
     & \mathcal{L}_{\text{SFT}}(\pi_{\text{ref}}, \mathcal{D}) = \mathcal{L}_{\text{ATT}}(\pi_{\text{ref}}, \mathcal{D}) + \alpha \cdot \mathcal{L}_{\text{CTC}}(\pi_{\text{ref}}, \mathcal{D}) , \\
     & \mathcal{L}_{\text{ATT}}(\pi_{\text{ref}}, \mathcal{D}) = - \sum_{u} \ln{P_{\text{ATT}}{(t_{u}^{\star}|a,v, t_{1:u-1}^{\star})}}, \\
     & \mathcal{L}_{\text{CTC}}(\pi_{\text{ref}}, \mathcal{D}) = - \ln{P_{\text{CTC}}(t^{\star}|a,v}),
\end{split}
\end{equation}
where $t_{1:u-1}^{\star}$ is the preceding tokens of the ground truth character sequence $y^{\star}$.
\begin{figure*}[t]
\centering
\includegraphics[width=0.9\linewidth]{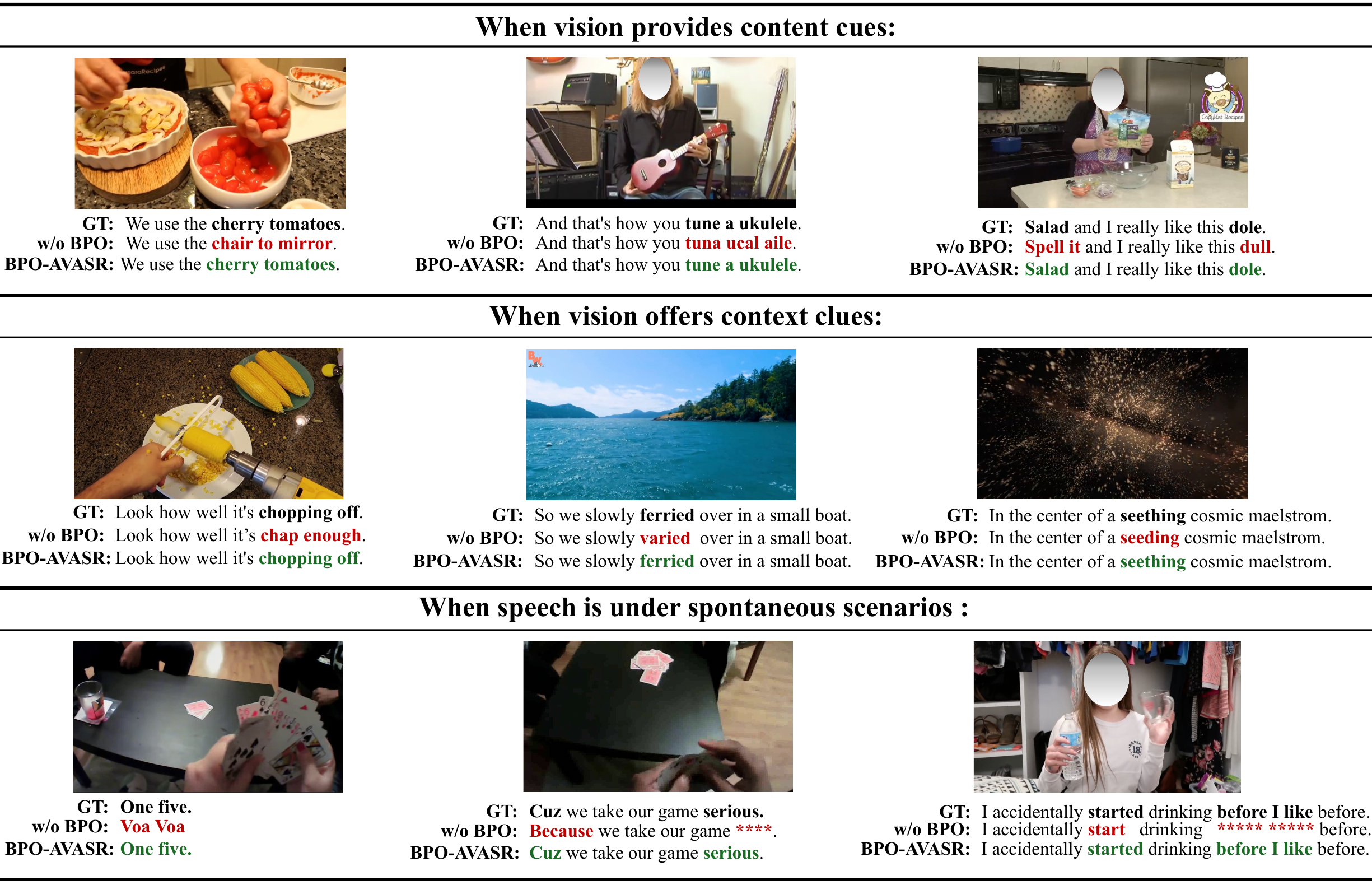} 
\caption{\textbf{Qualitative Results.} We show the ground truth text (GT), and predictions from the OWSM-visual small(w/o DPO) and BPO-AVASR small. We show the enhanced performance in three scenarios: when vision provides content cues (top), when
vision offers context clues (middle), and when speech is under spontaneous scenarios (bottom). Errors in the predicted words compared to the GT are highlighted in red. Faces are blurred for privacy.}
\label{fig:case_study}
\end{figure*}

\subsection{Bifocal Preference Optimization}
\label{sec:preference_opt}
To align the fine-tuned ASR model $\pi_{\text{ref}}$ (Equation~\ref{eq:SFT_loss}) with the AV-ASR task, we use input-side preference dataset $\widetilde{D}_c$ and the output-side preference dataset $\widetilde{D}_p$ (as described in Section~\ref{sec:dataset_construction}) to train the BPO-AVASR model $\pi_\theta$ through bifocal preference optimization.

Analogous to the Equation~\ref{eq:dpo_loss_1}, the bifocal preference optimization objective for AV-ASR is formulated as

\begin{equation}
\label{eq:dpo_loss}
\resizebox{0.45\textwidth}{!}{$
\begin{split}
  \mathcal{L}_{\text{BPO}}(\pi_\theta; \pi_\text{ref},\widetilde{D}) 
  & \! =\! \mathcal{L}_{\text{input}}(\pi_\theta; \pi_\text{ref}, \widetilde{D}_c) + \mathcal{L}_{\text{output}}(\pi_\theta; \pi_\text{ref}, \widetilde{D}_p),  \\
  \mathcal{L}_{\text{input}}(\pi_\theta; \pi_\text{ref}, \widetilde{D}_c) & \! = \! - \log \! \sigma \! \left( \beta \log \frac{\pi_{\theta}(t_w | c_w)}{\pi_{\text{ref}}(t_w|c_w)} 
  \! - \! \beta \log \frac{\pi_{\theta}(t_w|c_l)}{\pi_{\text{ref}}(t_w|c_l)} \right), \\
  \mathcal{L}_{\text{output}}(\pi_\theta; \pi_\text{ref}, \widetilde{D}_p) & \! = \! - \! \log \! \sigma \! \left( \beta \log \frac{\pi_{\theta}(t_w | c_w)}{\pi_{\text{ref}}(t_w|c_w)} 
  \! - \! \beta \log \frac{\pi_{\theta}(t_l|c_w)}{\pi_{\text{ref}}(t_l|c_w)} \right),  
\end{split}
$}
\end{equation}
where $\mathcal{L}_{\text{input}}$ and $\mathcal{L}_{\text{output}}$ are preference optimization loss for input-side preference and output-side preference respectively. $(t_l|c_w)$ and $(t_w|c_l)$ serve as hard negatives to $(t_w|c_w)$, sampled from the preference datasets $\widetilde{D}_p$ and $\widetilde{D}_c$ respectively. Through this approach, we train the BPO-AVASR model $\pi_\theta$ to avoid generating the above errors in the application, thereby
improving the ability of the AV-ASR model to take good advantage of complex audiovisual information, especially for unconstrained real-world videos.

\begin{table*}[t]
\centering
\begin{tabular}{lcccc}
\toprule
    \textbf{Model} & \textbf{Modality} & \textbf{How2} & \textbf{VisSpeech} &\textbf{ Ego4D} \\
    \midrule  
    OWSM-ft small~\cite{OWSM2023peng} & A & 10.8 & 16.6 & 70.6 \\
    BEST-RQ~\cite{Bestrq2022Chiu} & A & 15.3 & 16.7 & 68.3 \\
    \midrule
    How2 base~\cite{How22018Sanabria} & A+V & 18.0 & -- & -- \\
    VAT~\cite{VAT2019Caglayan} & A+V & 18.0 & -- & -- \\
    MultiRes~\cite{MultiRes2017Paraskevopoulos} & A+V  & 20.5 & -- & -- \\
    LLD~\cite{LLD2021Ghorbani} & A+V & 16.7 & -- & -- \\
    AVATAR~\cite{AVATAR2022Gabeur} & A+V & 15.6 & 43.4 & -- \\
    SynesLM~\cite{syneslm2024lu}& A+V & 15.7 & 39.4 & -- \\
    AVATAR$^\dagger$~\cite{AVATAR2022Gabeur} & A+V & \textbf{9.1} & 35.7 & 92.0 \\
    AVFormer$^\dagger$~\cite{AVFormer2023Seo} & A+V & 13.6 & 16.4 & 64.6 \\
    \midrule
    BPO-AVASR small & A+V & 9.3 (31.6\%) & \underline{15.6} (4.9\%) & \underline{59.2} (8.4\%) \\
    BPO-AVASR medium & A+V & \underline{9.2} (32.4\%) & \textbf{14.5} (11.6\%) & \textbf{56.5} (12.5\%) \\
    \bottomrule
    \end{tabular}

\caption{\textbf{Comparison to baseline methods across different datasets.} Results are reported as WER\%. For the VisSpeech and Ego4D datasets, models are evaluated without any fine-tuning. $^\dagger$ denotes these models are trained on a much larger additional audiovisual dataset, HowTo100M. ``A'' and ``V'' refer to audio and visual, respectively. Bolded values indicate the best results, while underlined values indicate the second-best results. The numbers in parentheses indicate the improvement in model performance (relative reduction in WER) compared to the state-of-the-art model considering three test sets, AVFormer$^\dagger$. }
\label{tab:overall}
\end{table*}

\section{Experimental Settings}
\subsection{Implementation Details}
As described in Section~\ref{sec:owsm}, we use OWSM v3.1 as our speech recognition backbone and develop two versions of BPO-AVASR: BPO-AVASR small and BPO-AVASR medium respectively. Specifically, BPO-AVASR-small consists of 9 E-Branchformer~\cite{Ebranchformer2022Kim} encoder blocks and 9 Transformer decoder blocks. The hidden size is 768. BPO-AVASR-medium comprises 18 E-Branchformer encoder blocks and 18 Transformer decoder blocks with a hidden size of 1024. Following AVFormer~\cite{AVFormer2023Seo}, we utilize CLIP-Large~\cite{clip2021Radford} as the visual encoder. For each video segment, we sample 4 frames uniformly (i.e. $M\!=\!4$ as described in Section~\ref{sec:SFT}). During the SFT stage, we fine-tune OWSM v3.1 for 10 epochs with a batch size of 64. We set $\alpha$ to 0.3 in Equation~\ref{eq:SFT_loss}.
In the BPO fine-tuning stage, we set $\beta$ to 0.1 in Equation~\ref{eq:dpo_loss}. We conduct the training using 1 V100 GPU, with a batch size of 512 and a learning rate of 2e-6. We use word error rate (WER) as the evaluation metric for all experiments, with lower values indicating better performance.

\subsection{Datasets}
In the supervised fine-tuning stage, we apply How2 as the fine-tuning dataset. In the bifocal preference optimization stage, we construct the preference dataset based on the How2. 
BPO-AVASR models are evaluated on three datasets, How2, VisSpeech, and Ego4D.

\begin{itemize}
    \item \textbf{How2}~\cite{How22018Sanabria} is an instructional video dataset designed for multimodal understanding. Following AVFormer, we use the 300-hour version of How2. These videos are segmented into short clips, averaging 5.8 seconds each, with a user-uploaded transcript. 
    
    \item \textbf{VisSpeech}~\cite{AVATAR2022Gabeur} is an AV-ASR benchmark sampled from the HowTo100M dataset~\cite{HowTo1002019Miech}. It contains 508 video clips with manually annotated transcripts. VisSpeech uses a video-text similarity model to ensure high audiovisual correspondence.
    
    \item \textbf{Ego4D}~\cite{Ego4D2022Grauman} is an egocentric daily-life activity video dataset. We use the audiovisual diarization benchmark and evaluate our model on the validation set with ground truth annotations.
    Videos are segmented into shorter clips for analysis. Unlike other datasets, Ego4D contains noisier and more spontaneous videos across different domains, increasing the difficulty of AV-ASR.

\end{itemize}

\subsection{Baselines}
We compare BPO-AVASR models to strong baselines, including robust audio-only ASR models (see the upper section of Table~\ref{tab:overall}) and state-of-the-art AV-ASR models (see the middle section of Table~\ref{tab:overall}).
\begin{itemize}
    \item \textbf{Audio-only ASR Models.} We compare BPO-AVASR series with (\romannumeral1) \textbf{BEST-RQ}, a robust ASR model pre-trained on LibriLight~\cite{Librilight2020Kahn}, LibriSpeech~\cite{LibriSpeech2015Panayotov} and HowTo100M~\cite{HowTo1002019Miech} datasets ($\sim$190K hours data in total); (\romannumeral2) \textbf{OWSM-ft small}, the OWSM v3.1 small model pre-trained on public speech datasets and fine-tuned on the How2 using audio-only data ($\sim$180K hours data in total).
    \item \textbf{AV-ASR Models.} We compare the BPO-AVASR series with the following AV-ASR baselines:
    \textbf{How2 base}~\cite{How22018Sanabria} re-trains an ASR model by learning a video-specific bias from speech features. \textbf{VAT}~\cite{VAT2019Caglayan} integrates different visual features into the ASR model with adaptive training. \textbf{MultiRes}~\cite{MultiRes2017Paraskevopoulos} fuses video features with an additional cross-modal multi-head attention layer. \textbf{LLD}~\cite{LLD2021Ghorbani} uses a deliberation model to leverage video and text representations extracted from a self-supervised text-video embedding model. \textbf{AVATAR}~\cite{AVATAR2022Gabeur} is an encoder-decoder based AV-ASR model which fuses visual and speech information through a multimodal encoder. \textbf{AVFormer}~\cite{AVFormer2023Seo} injects visual information into the frozen ASR model, BEST-RQ, with lightweight trainable adaptors, which shows good performance by training on 131k hours of audiovisual dataset. \textbf{SynesLM}~\cite{syneslm2024lu} is a unified speech language model capable of performing audiovisual speech recognition. It explores decoder-only architecture within a multitask learning objective. \textbf{OWSM-visual small} is the OWSM v3.1 small model fine-tuned on the How2 dataset that incorporates visual information (the first row of Table~\ref{tab:ablation_strategy}).
\end{itemize}

\begin{table*}[t]
    \centering
    \begin{tabular}{c|c|ccc}
    \toprule
        \multicolumn{2}{c|}{\textbf{Rejected Data Construction Strategy}} &   \textbf{How2} & \textbf{VisSpeech} & \textbf{Ego4D} \\
        \midrule 
        N/A (OWSM-visual small) & -- & 10.5 & 15.8  & 59.9 \\
        \midrule
        \multirow{3}{*}{Input-side preference} & Masked audio  & 9.4 & 15.7 & 59.2 \\
        \cmidrule{2-5}
        & Random cropped vision  & 12.1 & 15.7 & 61.5 \\
        & Flipped vision & 9.7 & 15.5 & 59.3 \\
        \midrule
        \multirow{4}{*}{Output-side preference} & Rule-based replacement& 10.4  & 16.0 & 62.4 \\
        & Homophone-based generation  & 9.5 & 15.8 & 59.8 \\
        & Spontaneous-based generation  & 10.8 & 16.1 & \textbf{59.0} \\
        & Vision-based generation & 9.7 & 15.6 & 59.3 \\       
        \midrule
        Mixture (BPO-AVASR small) & -- & \textbf{9.3} (11.4\%) & \textbf{15.6} (1.3\%) & 59.2 (1.7\%) \\
       
    \bottomrule
    \end{tabular}
    \caption{\textbf{Comparison of preference data constructing strategies}. Results are reported as WER\%. N/A indicates that preference data and preference optimization are not used. The digits in parentheses indicate the improvement in model performance (relative reduction in WER) compared to the model without preference optimization, i.e., OWSM-visual small.}
    \label{tab:ablation_strategy}
\end{table*}

\begin{table}[t]
    \centering
    \begin{tabular}{l|c}
    \toprule
        \textbf{Method}   & \textbf{WER\%} \\
        \midrule 
        AVATAR~\cite{AVATAR2022Gabeur} & 55.3 \\
        AVFormer~\cite{AVFormer2023Seo} & 55.2 \\
        OWSM-visual small  & 52.3 \\ 
        \midrule
        BPO-AVASR small & \textbf{50.0} (4.4\%) \\  
    \bottomrule
    \end{tabular}
    \caption{\textbf{Generalization to Ego4D dataset.} All models are fine-tuned on the Ego4D training set. 
    The digits in parentheses indicate the performance improvement (relative reduction in WER) compared to the OWSM-visual small.
    }
    \label{tab:ablation_ego4d}
\end{table}

\section{Experimental Results}
\subsection{Comparison with SOTA Models}
We compare BPO-AVASR with baseline models on three test sets in Table~\ref{tab:overall}. The BPO-AVASR models outperform all baselines trained on the same audiovisual dataset (How2 base, VAT, MultiRes, LLD, SynesLM, and AVATAR), particularly on the noisy and spontaneous Ego4D dataset, highlighting the effectiveness of bifocal preference optimization in adapting ASR to unconstrained AV-ASR tasks. Furthermore, BPO-AVASR medium achieves better results than BPO-AVASR small, demonstrating the benefit of using larger models.
While AVATAR$^\dagger$ achieves the best performance on How2, it shows poor generalization ability on the out-of-domain dataset Ego4D. In contrast, the BPO-AVASR series shows consistently better results across different domain datasets, demonstrating the robustness and generalization capability of BPO-AVASR. 
Compared with the previous state-of-the-art work AVFormer, the BPO-AVASR series shows significant improvement, especially on How2 (31.6\% and 32.4\%, respectively). 
Notably, both AVFormer$^\dagger$ and BPO-AVASR are based on pre-trained ASR models with over 100K hours of training data. However, while AVFormer$^\dagger$ is further fine-tuned on 131k hours of audiovisual data from HowTo100M, BPO-AVASR achieves superior results using only 300 hours of audiovisual data.

\subsection{Qualitative Analysis}
Examples in Figure~\ref{fig:case_study} illustrate the effectiveness of BPO-AVASR. Qualitative results show that OWSM-visual (w/o BPO) still struggles to obtain accurate transcripts in real-world video scenarios. In contrast, qualitative examples demonstrate how BPO-AVASR improves ASR performance. 
As shown in Figure~\ref{fig:case_study}, BPO-AVASR is effective in three scenarios: when vision provides content cues, when vision offers context clues, and when speech is under spontaneous scenarios.
For instance, visual content helps the model recognize objects directly in the video, such as brand names (row 1 column 3) and object names (row 1 column 1, row 1 column 2).
Additionally, by using visual information as contextual clues, the model can correctly distinguish between similarly pronounced words (row 2). 
Furthermore, by constructing rejected samples considering spontaneous words, BPO-AVASR improves the accuracy of recognizing filler words in spontaneous speech scenarios (row 3).

\subsection{Ablation Studies}
\paragraph{Analysis of Rejected Samples.}
\label{sec:ablation_strategy}
To evaluate the effects of different preference construction strategies, we create preference datasets with eight different strategies and then use them to optimize the BPO-AVASR small model. From the results in Table~\ref{tab:ablation_strategy}, we have the following observations:
\begin{itemize}
    \item \textbf{Overall.} The bifocal preference dataset constructed by mixing all strategies leads to the best and most balanced performance across all test sets. Moreover, when comparing BPO-AVASR small with OWSM-visual, the significant improvement (especially the 11.4\% improvement on How2) highlights the effectiveness of our proposed bifocal preference optimization over SFT optimization. 
    \item \textbf{Importance of Hard Negatives.} Constructing hard negative samples is crucial for providing effective preference optimization signals. For input-side preference, frame flipping shows better results than random cropping.
    We attribute this to the fact that random cropping often fails to remove useful information due to the redundancy of visual data in unconstrained AV-ASR, making it difficult to create truly hard negatives.
    For output-side preference construction, we compare the rule-based construction strategy with the LLM-based strategy described in Section~\ref{sec:dataset_construction}. Specifically, the rule-based strategy uses a homophone dictionary to replace content words in the ground truth transcripts. The LLM-based preference construction shows greater improvement, due to the LLM's strong text understanding and generation capabilities, which are more effective in constructing hard negatives.
    \item  \textbf{Effectiveness of Different Strategies.} Preference optimization strategies show different improvements on three test sets.
    Masking audio improves speech recognition performance across all three datasets. The Ego4D dataset benefits significantly from the strategy that addresses spontaneous speech (from 59.9 to 59.0).
\end{itemize}

\paragraph{Generalization to Other Datasets.}
To demonstrate the generalization ability of BPO-AVASR, we further construct a bifocal preference dataset using the Ego4D training set~\cite{Ego4D2022Grauman}, following the same strategies described in Section~\ref{sec:dataset_construction}. We then fine-tune BPO-AVASR small on this constructed preference dataset. As shown in Table~\ref{tab:ablation_ego4d}, BPO-AVASR outperforms all previous works, demonstrating the effectiveness of bifocal preference optimization on real-world videos, including instructional (How2) and egocentric (Ego4D) videos.
Moreover, fine-tuning on Ego4D with preference optimization results in a significant improvement in WER compared to zero-shot testing on Ego4D (Table~\ref{tab:overall}). This highlights that constructing preferences within the same domain further enhances the model’s performance through preference optimization.

\section{Conclusion}
In this paper, we first formulate the AV-ASR task as a preference optimization problem.
Accordingly, we develop BPO-AVASR, an AV-ASR system optimized by bifocal preference optimization to improve speech recognition accuracy for real-world videos. First, we introduce a simple yet effective method to create preference data by simulating common errors related to different modalities in AV-ASR. Second, we propose a bifocal preference optimization strategy to optimize AV-ASR models by emphasizing the distinction between correct and incorrect answers. BPO-AVASR outperforms previous state-of-the-art models, demonstrating the effectiveness of using preference optimization to align the audio-only ASR model to real-world video scenarios.
For future work, we plan to build a high-quality open-domain AV-ASR dataset to facilitate future research.
\bigskip

\section{Acknowledgments}
This work is supported by the National Natural Science Foundation of China (No. 62276268) and the Outstanding Innovative Talents Cultivation Funded Programs 2022 of Renmin University of China.
Experiments of this work used the Bridges2 system at PSC and Delta system at NCSA through allocations CIS210014 and IRI120008P from the Advanced Cyberinfrastructure Coordination Ecosystem: Services \& Support (ACCESS) program.

\bibliography{aaai25}
\newpage
% \clearpage
\begin{figure*}[t]
\centering
\includegraphics[width=\linewidth]{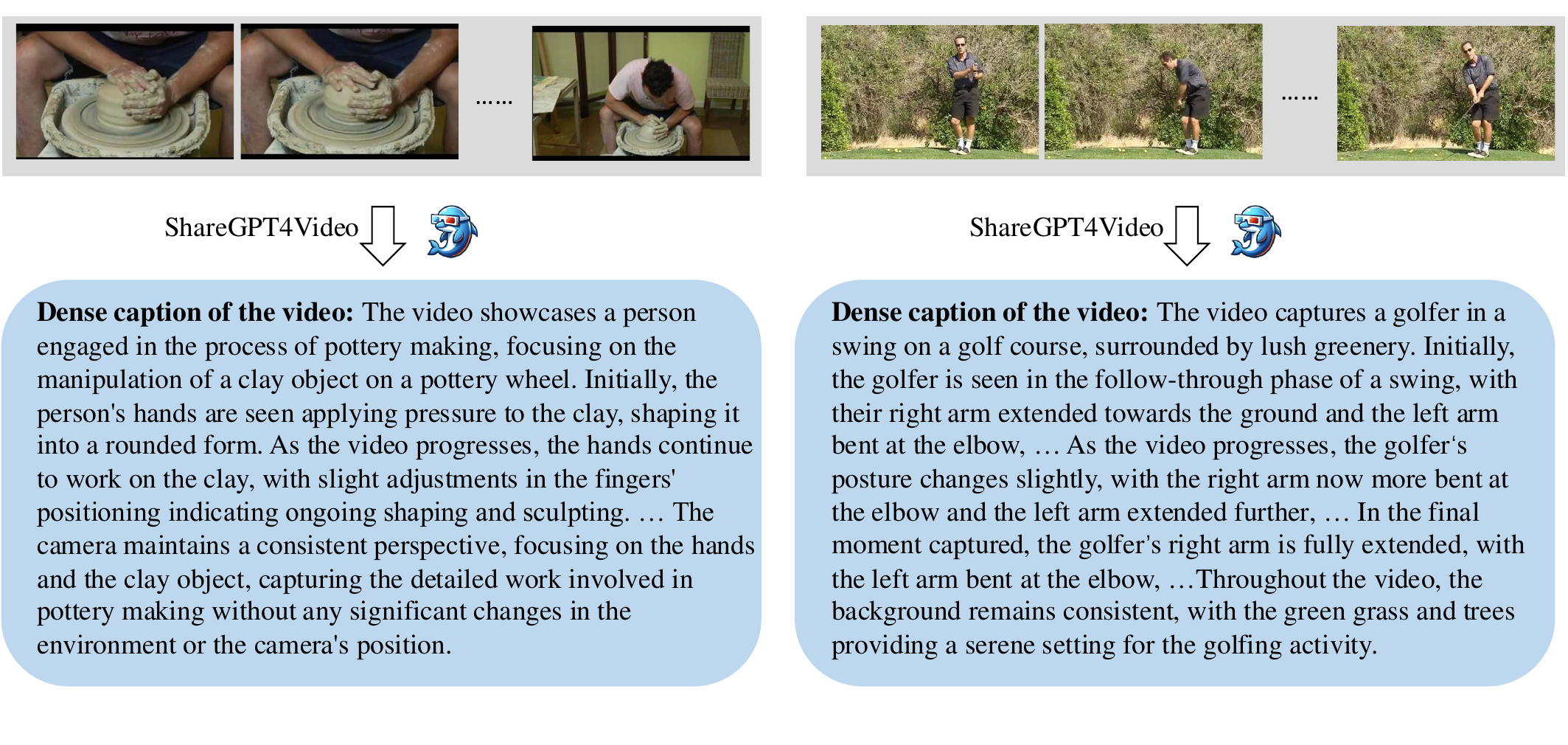} 
\caption{\textbf{Examples of generated dense captions using ShareCaptioner-Video.} }
\label{fig:dense_caption}
\end{figure*}
\section{Supplementary Material}
\subsection{Generation of Dense Caption}
As mentioned in Section 3.1, we employ ShareGPT4Video~\cite{chen2024sharegpt4video} to convert video streams to dense captions. More specifically, we use ShareCaptioner-Video, an efficient and capable captioning model for arbitrary videos. Figure~\ref{fig:dense_caption} shows examples of dense captions generated by ShareCaptioner-Video. Dense captions describe details of the input video streams, including inter-frame target actions and behaviors, changes in environment and background, alterations in target appearance attributes, and camera movements reflecting temporal changes. This comprehensive information helps in generating vision-related rejected samples by masking specific visual information during the output-side preference construction.

\subsection{Prompt of Output-side Preference Construction}
As shown in Table~\ref{tab:prompts}, we use ChatGPT to generate output-side preference. By designing different prompts, we manipulate the transcripts with homophone-based generation, spontaneous-based generation, and vision-based generation. We will release our constructed preference dataset upon acceptance.

\begin{table*}
\centering
\begin{tabular}{l|l|l}
\toprule
 \textbf{Strategy and ground truth transcript} & \textbf{Prompt to ChatGPT} & \textbf{Output from ChatGPT} \\
\midrule  
\makecell[l]{Type \uppercase\expandafter{\romannumeral4}: Spontaneous-based Generation \\ \textbf{transcript\textsubscript{GT}}: \textit{We use marble bases}\\ \textit{that are imported directly from Italy.}} & \makecell[l]{We have a speech for which the original\\ transcription is \textbf{transcript\textsubscript{GT}}. Can you \\generate a candidate transcription that \\ includes common ASR mistakes? Specifically, \\recognize or omit spontaneous words incorrectly.} & \makecell[l]{\textit{We use marble bases, like,} \\ \textit{imported directly from Italy.}}\\
\midrule

    \makecell[l]{Type \uppercase\expandafter{\romannumeral3}: Homophone-based Generation \\ \textbf{transcript\textsubscript{GT}}: \textit{We use marble bases} \\ \textit{that are imported directly from Italy.}} & \makecell[l]{We have a speech for which the original \\ transcription is \textbf{transcript\textsubscript{GT}}. Can you \\ generate a candidate transcription that \\ includes common ASR mistakes? Specifically, \\ generate words with similar pronunciation,\\ especially for content words} & \makecell[l]{\textit{We use marble vases that are} \\ \textit{imported directly from Italy.}}\\
    \midrule
    
    \makecell[l]{Type \uppercase\expandafter{\romannumeral5}: Vision-based Generation \\ \textbf{transcript\textsubscript{GT}}: \textit{We use marble bases}\\ \textit{that are imported directly from Italy.}} & \makecell[l]{We have a speech for which the original\\ transcription is \textbf{transcript\textsubscript{GT}}. This speech is\\ from a video, the dense caption of this video \\ is \textbf{caption\textsubscript{dense}}. Can you generate a candidate \\transcription that includes common ASR \\ mistakes? Specifically, generate words that\\ ignore the video content, especially for words\\ that appear or are described in the video caption.} & \makecell[l]{\textit{We use mobile bases that are} \\ \textit{imported directly from Italy.}}\\

\bottomrule
\end{tabular}
\caption{\textbf{Prompts used in ChatGPT for creating the output-side preference.} transcript\textsubscript{GT} refers to the ground truth transcripts, while caption\textsubscript{dense} refers to the dense caption of the video.}
\label{tab:prompts}
\end{table*}

% Bibliography

%Manual citation list
%\begin{thebibliography}{1}
%\bibitem{Zhang:14}
%Y.~Zhang, S.~Qiao, L.~Sun, Q.~W. Shi, W.~Huang, %L.~Li, and Z.~Yang,
 % \enquote{Photoinduced active terahertz metamaterials with nanostructured
  %vanadium dioxide film deposited by sol-gel method,} Opt. Express \textbf{22},
  %11070--11078 (2014).
%\end{thebibliography}

\end{document}